\documentclass[aps,pra,preprint,showpacs,showkeys,nofootinbib]{revtex4-1}
\usepackage{eurosym}
\usepackage{amsfonts}
\usepackage{amsmath}
\usepackage{amssymb}
\usepackage{graphicx}
\usepackage{epsfig}
\usepackage{color}

\providecommand{\U}[1]{\protect\rule{.1in}{.1in}}

\begin{document}

\preprint{}
\title{High-order harmonic generation driven by metal nanotip photoemission: theory and simulations}

\author{M. F. Ciappina$^{1}$}
\author{J. A. P\'erez-Hern\'andez$^{2}$}
\author{M. Lewenstein$^{3,4}$}
\author{M. Kr\"uger$^{5}$}
\author{P. Hommelhoff$^{5}$}

\affiliation{$^{1}$Department of Physics, Auburn University, Auburn, Alabama 36849, USA}
\affiliation{$^{2}$Centro de L\'aseres Pulsados (CLPU), Parque Cient\'{\i}fico, 37185 Villamayor, Salamanca, Spain}
\affiliation{$^{3}$ICFO-Institut de Ci\'encies Fot\'oniques, Mediterranean Technology
Park, 08860 Castelldefels (Barcelona), Spain}
\affiliation{$^{4}$ICREA-Instituci\'o Catalana de Recerca i Estudis Avan\c{c}ats, Lluis
Companys 23, 08010 Barcelona, Spain}
\affiliation{$^{5}$ Department of Physics, Friedrich-Alexander-Universit\"at Erlangen-N\"urnberg, Staudtstr.~1, D-91058 Erlangen, Germany, and Ultrafast Quantum Optics Group, Max-Planck-Institut f\"ur Quantenoptik, Hans-Kopfermann-Str.~1, D-85748 Garching bei M\"unchen, Germany}


\begin{abstract}
We present theoretical predictions of high-order harmonic generation (HHG) resulting from the interaction of short femtosecond laser pulses with metal nanotips. It has been demonstrated that high energy electrons can be generated using nanotips as sources; furthermore the recollision mechanism has been proven to be the physical mechanism behind this photoemission. If recollision exists, it should be possible to convert the laser-gained energy by the electron in the continuum in a high energy photon. Consequently the emission of harmonic radiation appears to be viable, although it has not been experimentally demonstrated hitherto. We employ a quantum mechanical time dependent approach to model the electron dipole moment including both the laser experimental conditions and the bulk matter properties. The use of metal tips shall pave a new way of generating coherent XUV light with a femtosecond laser field.
\end{abstract}

\maketitle

One of the most prominent examples of the nonlinear interaction between laser and matter, atoms and molecules, is high-order harmonic generation (HHG) process~\cite{McPherson1987,Huillier1991}. The great interest of this phenomenon resides in the fact that HHG represents one of the most reliable pathways to generate coherent ultraviolet (UV) to extreme ultraviolet (XUV) light. In addition, HHG based on atoms and molecules has proven to be a robust source for the generation of attosecond pulses trains~\cite{corkumnat}, that can be temporally confined to a single XUV attosecond pulse, now with high repetition rates~\cite{Scrinzi2004,hhg1khz,hhg3khz,amelle}. Thanks to its remarkable properties, HHG can as well be employed to extract temporal and spatial information with both attosecond and sub-{\AA}ngstr\"om resolution on the generating system~\cite{manfred_rev}. Furthermore, it represents a considerable tool to scrutinize the atomic world within its natural temporal and spatial scales~\cite{pacer,rabitt,olga1,olga2,mairesse,power}.

Instead of using atoms and molecules in the gas phase as active medium, the utilization of bulk matter, for instance metal nanotips~\cite{peter2011}, nanoparticles~\cite{klingnature} or ablation plumes~\cite{ganeev1,ganeev2}, has recently been put forward (see e.g.~\cite{peter2012} for more references). For instance, the laser matter phenomenon called above-threshold photoemission (ATP) represents the counterpart of the above-threshold ionization (ATI) in atoms and molecules,
but the underlying physics is much richer and quite different in nature (see e.g.~\cite{peter2010,herink} for field-enhancement and near-field effects). Several models have been recently developed and applied in order to understand the experimental data and to guide future measurements~\cite{yalunin,michaelNJP,watcherPRB,ropersAdP}.  Furthermore, it has been demonstrated that solid state samples can also be used as a generators of high order harmonic radiation, although the HHG phenomenon using bulk matter is at its very beginning, both theoretically and experimentally~\cite{ghimireexp,ghimiretheory}.

Another related process, combining noble gases and bulk matter, is the generation of harmonic radiation using (plasmonically) enhanced fields. The first demonstration of such an effect was obtained employing surface plasmonic resonances that can locally amplify the incoming laser field~\cite{kim}. Using such resonances, amplifications in intensity greater than 20 dB can be achieved~\cite{muhl, schuck}. Consequently, when a low intensity femtosecond laser pulse couples to the plasmonic mode of a metal nanoparticle, it initiates a collective oscillation among free electrons within the metal. A region of highly amplified electric field, exceeding the threshold for HHG, can so be generated while these free charges redistribute the response field around the metal nanostructure. By injection of noble gases surrounding the nanoparticle high order harmonics can be produced. Particularly, using gold bow-tie shaped nanostructures it has been demonstrated that the initially modest laser field can be amplified sufficiently to generate high energy photons, in the XUV regime, and the radiation generated from the enhanced laser field, localized at each nanostructure, acts as a point-like source, enabling collimation of this coherent radiation by means of constructive interference~\cite{kim}. Recently there has been extensive theoretical work looking at HHG driven by spatially nonhomogeneous fields~\cite{husakou,yavuz,ciappi2012,tahir2012,ciappi_opt,ciappiAdP,miloAdP,joseprl2013,yavuz2013,luo2013}. However, the initial sudden excitement about the utilization of plasmonic fields for HHG in the XUV range was put in debate by recent findings~\cite{ropersnat, Kimreply, sivis2013}. Fortunately, alternative ways to enhance coherent light were explored (e.g. the production of high energy photoelectrons using enhanced near-fields from dielectric nanoparticles~\cite{klingnature}, metal nanoparticles~\cite{klingspie,klingprb,lastkling} and metal nanotips~\cite{herink,peter2010,peter2011,peter2006,peter2006a,Ropers2007,Barwick2007,Yanagisawa2009,Bormann2010,Park2012,peter2012}.

In this contribution we predict that it is entirely possible to generate high-order harmonics directly from metal nanotips. We employ available laser source parameters and model the metal tip with a fully quantum mechanical model. As it is well known, the main physical mechanism behind the generation of high order harmonics is the electron recollision and consequently the model used should include it. It was already shown that the recollision mechanism is also needed to describe  above-threshold photoemission (ATP) measurements and, considering these two laser-matter phenomena, i.e. the photoemitted electrons and the high frequency radiation, are physically linked, we could argue that metal nanotips can be used as sources of XUV radiation as well.

The theoretical model we use here has already been described elsewhere and employed for the calculation of the electron photoemission in metal nanotips~\cite{peter2006a,peter2012}.
As a consequence we only give a brief overview and we emphasize the numerical tools needed to compute the HHG spectra.
In short, the one dimensional time dependent Schr\"odinger equation (1D-TDSE) is solved for a single active electron
in a model potential. We employ a narrow, few atomic units wide, potential well with variable depth ($W+E_F,$ where $W$ is the work function and $E_F$ the Fermi energy) to model the metal surface.
This depth and width of the well are chosen in such a way as to match the actual metal tip parameters. In our case we
employ the parameters for clean gold, i.e. $W=5.5$ eV and a $E_F=4.5$ eV, but other typical metals, as tungsten, can be used as well. The ground
state of the active electron represents the initial state in the metal nanotip. The electronic wavefunction is confined by
an infinitely high potential wall on one side and on the other
side by a potential step representing the metal-vacuum surface
barrier. In addition we consider an image-force potential that gives a smoother shape to the surface barrier potential.
The evanescent part of the electronic wavefunction penetrates into
the classically forbidden (vacuum) region. The rescattering mechanism, mainly responsible of the high energy region of the photoelectron spectra and
the high order harmonic generation, is closely linked with this evanescent behavior.

We employ laser pulses of the form $E_{L}(t)=E_0\,f(t)\sin(\omega t+\phi_{CEP})$, with $E_0$, $\omega$ and $\phi_{CEP}$ are the laser electric field peak amplitude, the laser frequency and the carrier envelope phase (CEP), respectively. The envelope $f(t)$ can be chosen between sine-squared or trapezoidal shape. The former allows us to model ultrashort laser pulses, 2-4 cycles long, and to study the effects of the $\phi_{CEP}$ in the HHG spectra. On the other hand, the latter is needed in order to model long pulses. In our computations we employ a pulse 10 full cycles long with 2 cycles of turn on and off and 6 cycles of constant amplitude. In addition to the laser
electric field, we include a static field $E_{dc}$ that arises
due to the tip bias voltage. As a consequence we can write the potential the electron feels, without considering the image force, as $V(z,t)=(E_{dc}+E_{L}(t))z$. The electronic wavefunction is time propagated using the Crank-Nicolson scheme under the influence of the external fields. Finally, the harmonic spectra are retrieved by Fourier-transforming the dipole acceleration that is obtained from the electronic dipole moment (for details see e.g.~\cite{keitel,schafer}).

%\begin{figure}[htb]
%\centering
%%\includegraphics[width=0.3\textwidth]{Figure3}
%\caption{Idem Fig.~1 but now $I=5\times10^{13}$ W/cm$^{2}$.}
%\label{fig:figure3}
%\end{figure}

%\begin{figure}[htb]
%\centering
%%\hspace{-3cm}
%\includegraphics[width=0.7\textwidth]{Figure3}
%%\vspace{-1cm}
%\caption{(color online) HHG spectra as a function of harmonic order for a metal (Au) nanotip using a trapezoidal shaped laser pulse with 10 cycles of total time and a wavelength $\protect\lambda=685$ nm. Panel a) $E_0=10$ GV m$^{-1}$; panel b)  $E_0=15$ GV m$^{-1}$ and panel c)  $E_0=20$ GV m$^{-1}$. In all the panels magenta: $E_{dc}=-0.4$ GV m$^{-1}$, green: $E_{dc}=+2$ GV m$^{-1}$.}
%\label{fig:figure3}
%\end{figure}

%\begin{figure}[htb]
%\centering
%%\hspace{-3cm}
%\includegraphics[width=0.7\textwidth]{Figure4}
%%\vspace{-1cm}
%\caption{(color online) HHG spectra as a function of harmonic order for a metal (Au) nanotip using a trapezoidal shaped laser pulse with 10 cycles of total time and a wavelength $\protect\lambda=1800$ nm and a peak laser electric field $E_0=10$ GV m$^{-1}$. Magenta $E_{dc}=-0.4$ GV m$^{-1}$, green $E_{dc}=+2$ GV m$^{-1}$.}
%\label{fig:figure4}
%\end{figure}

%\begin{figure}[htb]
%\centering
%%\includegraphics[width=0.4\textwidth]{Figure4}
%\caption{(color online) Dependence of the semiclassical trajectories on the
%ionization ($t_i$) and recollision ($t_r$) times for different values of $\chi$. Red (dark gray) squares are homogeneous case, i.e. $\chi\rightarrow\infty$; green (light gray) circles are $\protect\chi=50$ and black triangles $\chi=40$.}
%\label{fig:figure4}
%\end{figure}

We first study the CEP dependence of the harmonic spectra generated when a metal (Au) nanotip is illuminated by a femtosecond laser pulse. The experimental confirmation of our predictions could be done by employing a similar experimental set-up as the one used for electron photoemission, although some questions remain to be solved~\cite{peter2006,peter2011,peter2012}. For instance, we are unable in our calculations to estimate absolute values for the harmonic signal. In addition, we model a single atom response. Propagation and mode-matching effects are completely neglected in our approach. These last could play a role, although it has been argued in two recent experiments~\cite{kim,funnel} that in the harmonic emission from nanosources one could safely ignore them due to the strong confinement of the radiation sources in dimensions smaller than the laser wavelength.

%
%{\bf (PH: from here on: discuss this later? Also add: no mode-matching (propagation effects): a much less collimated beam of XUV light expected; mention bulk effects? Could it be possible that considering the dipole moment is not the whole story? Some plasma effect to include? Brunel effect (PRL Vol. 59, p. 52)??) The experimental confirmation of our predictions could be done by employing a similar experimental set-up as the one used for electron photoemission, although some challenges would have to be solved~\cite{peter2006,peter2011,peter2012}. For instance, we are unable in our calculations to estimate absolute values for the harmonic signal. In addition, propagation and mode-matching effects are completely neglected in our approach and this could play a role, although it has been argued in two recent experiments~\cite{kim,funnel} that in the harmonic emission from nanosources one could safely ignore them due to the strong confinement of the radiation sources.}. 

We use ultrashort laser pulses with a full width at half maximum (FWHM) duration of 2 fs and 4 fs (corresponding approximately to a 2 and 4 total cycles), wavelength $\lambda=800$ nm (photon energy 1.55 eV) and we vary the carrier envelope phase $\phi_{CEP}$ in order to cover its full range ($[0,2\pi]$). 
We model laser peak fields of up to 20 GV m$^{-1}$, as this is close to the
experimentally accessible range before damage may set in. Furthermore, plasma effects do not appear whenever the laser intensity does not exceed the saturation limit of each specific chemical specimen target (gold, in this case) and for the mentioned values we are below such limit. Note that the
values of the peak electric fields mentioned here mean the physical fields at the tip, including the
effects of field enhancement. For the static field $E_{dc}$ we employ a typical value, $E_{dc}=-0.4$ GV m$^{-1}$ and also $E_{dc}=+2$ GV/m, corresponding to a positive tip bias voltage. 

In Fig.~1 we show calculations of the harmonic yield as a function of the harmonic order and $\phi_{CEP}$ for the 2 fs FWHM laser pulse case; Fig.~2 presents the 4 fs FWHM laser pulse case. The different panels correspond to a peak electric field $E_0=10$ GV m$^{-1}$ and a static field $E_{dc}=-0.4$ GV m$^{-1}$ (Fig.~1(a)); $E_0=10$ GV m$^{-1}$ and $E_{dc}=+2$ GV m$^{-1}$ (Fig.~2(a)); $E_0=20$ GV m$^{-1}$ and $E_{dc}=-0.4$ GV m$^{-1}$  (Fig.~2(a)) and $E_0=20$ GV m$^{-1}$ and  $E_{dc}=+2$ GV m$^{-1}$  (Fig.~2(b)).
The first feature we can observe is the strong modulation in the spectra as the $\phi_{CEP}$ changes, more pronounced for the 2 fs FWHM case (similar behavior is observed in atoms, see e.g.~\cite{tanos,amelle}). A clear harmonic cutoff can be seen for the 2 fs FWHM case at $n_c\approx5$ (equivalent to a photon energy of 7.75 eV) for $E_0=10$ GV m$^{-1}$ (Fig.~1(a) and Fig.~1(b)) and one at $n_c\approx 10$ (equivalent to a photon energy of 15.5 eV) for $E_0=20$ GV m$^{-1}$ (Fig.~2(a) and Fig.~2(b)), although the latter is more visible for particular values of the $\phi_{CEP}$. When we increase the pulse length to 4 fs FWHM the modulation is less noticeable, but nevertheless harmonic cutoff values similar to the previous cases are observed. For $E_{dc}>0$ an increase in the harmonic yield is observed for some values of $\phi_{CEP}$ and it is more clear for the cases when $E_0=20$ GV m$^{-1}$. This feature could be experimentally exploited as a larger signal will be easier to detect.

We can consider the semiclassical simple man's model predictions~\cite{corkum,sfa} in order to characterize the harmonic cutoff, although the influence of the static field $E_{dc}$ will not be included. It is well established that for atoms and molecules $n_{c}=(3.17U_{p}+I_{p})/\omega$, where $n_{c}$ is the harmonic order at the cutoff and $U_{p}$ the ponderomotive energy ($U_p=E_0^{2}/4\omega^2$)~\cite{keitel} and $I_p$ the ionization potential of the atomic or molecular species under consideration. Inserting the values of the peak electric field and laser wavelength and using an equivalent $I_p$ equal to the metal work function $W$ we can corroborate the cutoff values obtained from our quantum mechanical model. For instance, for $E_0=10$ GV m$^{-1}$ (0.02 a.u) $n_c\approx 5$ and this value is in very good agreement with the quantum mechanical calculations (see. e.g. Fig.~1(a), Fig.~1(b), Fig.~2(a) and Fig.~2(b)). For positive values of $E_{dc}$ it appears that the harmonic cutoff increases, but the effect could be masked by the CEP influence. For longer pulses, however, we do observe a clear harmonic extension for $E_{dc}>0$ (see below).

Next, in Fig.~3, we show harmonic spectra by using a long (10 cycles) trapezoidally shaped laser pulse of $\lambda=685$ nm (photon energy 1.81 eV). The different panels correspond to various values of the peak electric field $E_0$, namely $10$ GV m$^{-1}$, $15$ GV m$^{-1}$ and $20$ GV m$^{-1}$ for Fig.~3(a), Fig.~3(b) and Fig.~3(c), respectively. In the three panels the magenta/dark gray (green/light gray) represent the case of $E_{dc}=-0.4$ GV m$^{-1}$ ($E_{dc}=+2$ GV m$^{-1}$). We can observe an increasing of the relative yield in the plateau region for positive values of the $E_{dc}$ field. This gain in conversion efficiency is again important for ease of experimental radiation detection.

Finally we compute in Fig.~4 harmonic spectra by using a long (10 cycles) trapezoidal shaped laser pulse of $\lambda=1800$ nm (photon energy 0.69 eV) and a peak electric field $E_0=10$ GV m$^{-1}$. The magenta/dark gray curve is for $E_{dc}=-0.4$ GV m$^{-1}$ while the green/light gray one is for $E_{dc}=+2$ GV m$^{-1}$. We can observe harmonic cutoffs of around $n_c=30$ (that correspond to an equivalent energy of 20 eV) for the former case and $n_c=60$ (41 eV) for the latter. In here, the semiclassical model predicts $n_c\approx 27$ and we could observe that the influence of the static electric field appears to be more pronounced when $\lambda$ increases. Considering the semiclassical cutoff is proportional to $\lambda^2$ it is evident that the utilization of longer wavelengths laser sources would allow us to reach high energy photons.

In conclusion, we predict that it is possible to generate high order harmonics directly from metal nanotips. We employ a quantum mechanical model in order to compute the HHG yield and we use typical laser parameters available experimentally. A comparable approach was successfully applied to predict the photoelectron spectra under similar experimental conditions~\cite{peter2012}. We observe a strong modulation of the HHG spectra with the variation of the $\phi_{CEP}$ for short pulses and a noticeable extension of the HHG cutoff when negative static fields are employed, more pronounced when longer wavelengths laser sources are employed. For the parameters we have used we can safely neglect any spatial variation of the (plasmonically) enhanced field. For instance, using $\lambda=1800$ nm and $E_0=10$ GV m$^{-1}$ (0.02 a.u.) the classical electron quiver radius $\alpha$  ($\alpha=E_0/\omega^{2}$) is around 1.6 nm and, for typical metal nanotips with radii between $R=5$ nm and $R=50$ nm, the estimated optical field decay length (1/e) is around $L=(0.82\pm0.04) R$~\cite{thomas}. Consequently the laser-ionized electron would practically feel a spatially homogeneous electric field as $\alpha\ll L$ for the parameters given. The aspect of the spatially inhomogenuous field will become important for other experimental conditions, which will be considered in a future work. In addition, we plan to extend the simple man's model (SMM), already employed for electron emission in metal nanotips, to treat HHG. In this way, a deeper understanding of the underlying physics of harmonic emission from metal nanotips will be achieved.

We acknowledge the financial support of the MICINN
projects (FIS2008-00784 TOQATA, FIS2008-06368-C02-01,
and FIS2010-12834), ERC Advanced Grant QUAGATUA, the
Alexander von Humboldt Foundation, the Hamburg Theory
Prize (M.L.), and the DFG Cluster of Excellence Munich Center for Advanced Photonics. This research has been partially supported
by Fundaci\`o Privada Cellex. J.A.P.-H. acknowledges support
from the Spanish MINECO through the Consolider Program
SAUUL (CSD2007-00013) and research project FIS2009-
09522, from Junta de Castilla y Le\'on through the Program
for Groups of Excellence (GR27), and from the ERC Seventh
Framework Programme (LASERLAB-EUROPE, Grant No.
228334). This work was made possible in part by a grant of high performance computing
resources and technical support from the Alabama Supercomputer Authority.

%\end{document}

%\bibliographystyle{plain}

%\bibliography{plasmonics}

\begin{thebibliography}{60}
\expandafter\ifx\csname natexlab\endcsname\relax\def\natexlab#1{#1}\fi
\expandafter\ifx\csname bibnamefont\endcsname\relax
  \def\bibnamefont#1{#1}\fi
\expandafter\ifx\csname bibfnamefont\endcsname\relax
  \def\bibfnamefont#1{#1}\fi
\expandafter\ifx\csname citenamefont\endcsname\relax
  \def\citenamefont#1{#1}\fi
\expandafter\ifx\csname url\endcsname\relax
  \def\url#1{\texttt{#1}}\fi
\expandafter\ifx\csname urlprefix\endcsname\relax\def\urlprefix{URL }\fi
\providecommand{\bibinfo}[2]{#2}
\providecommand{\eprint}[2][]{\url{#2}}

\bibitem[{\citenamefont{McPherson et~al.}(1987)\citenamefont{McPherson, Gibson,
  Jara, Johann, Luk, McIntyre, Boyer, and Rhodes}}]{McPherson1987}
\bibinfo{author}{\bibfnamefont{A.}~\bibnamefont{McPherson}},
  \bibinfo{author}{\bibfnamefont{G.}~\bibnamefont{Gibson}},
  \bibinfo{author}{\bibfnamefont{H.}~\bibnamefont{Jara}},
  \bibinfo{author}{\bibfnamefont{U.}~\bibnamefont{Johann}},
  \bibinfo{author}{\bibfnamefont{T.~S.} \bibnamefont{Luk}},
  \bibinfo{author}{\bibfnamefont{I.~A.} \bibnamefont{McIntyre}},
  \bibinfo{author}{\bibfnamefont{K.}~\bibnamefont{Boyer}}, \bibnamefont{and}
  \bibinfo{author}{\bibfnamefont{C.~K.} \bibnamefont{Rhodes}},
  \bibinfo{journal}{J. Opt. Soc. Am. B} \textbf{\bibinfo{volume}{4}},
  \bibinfo{pages}{595} (\bibinfo{year}{1987}).

\bibitem[{\citenamefont{L'Huillier et~al.}(1991)\citenamefont{L'Huillier,
  Schafer, and Kulander}}]{Huillier1991}
\bibinfo{author}{\bibfnamefont{A.}~\bibnamefont{L'Huillier}},
  \bibinfo{author}{\bibfnamefont{K.~J.} \bibnamefont{Schafer}},
  \bibnamefont{and} \bibinfo{author}{\bibfnamefont{K.~C.}
  \bibnamefont{Kulander}}, \bibinfo{journal}{J. Phys. B}
  \textbf{\bibinfo{volume}{24}}, \bibinfo{pages}{3315} (\bibinfo{year}{1991}).

\bibitem[{\citenamefont{Corkum and Krausz}(2007)}]{corkumnat}
\bibinfo{author}{\bibfnamefont{P.~B.} \bibnamefont{Corkum}} \bibnamefont{and}
  \bibinfo{author}{\bibfnamefont{F.}~\bibnamefont{Krausz}},
  \bibinfo{journal}{Nat. Phys.} \textbf{\bibinfo{volume}{3}},
  \bibinfo{pages}{381} (\bibinfo{year}{2007}).

\bibitem[{\citenamefont{Scrinzi et~al.}(2004)\citenamefont{Scrinzi,
  Westerwalbesloh, Kleineberg, U.Heinzmann, Drescher, and
  Krausz}}]{Scrinzi2004}
\bibinfo{author}{\bibfnamefont{A.}~\bibnamefont{Scrinzi}},
  \bibinfo{author}{\bibfnamefont{T.}~\bibnamefont{Westerwalbesloh}},
  \bibinfo{author}{\bibfnamefont{U.}~\bibnamefont{Kleineberg}},
  \bibinfo{author}{\bibnamefont{U.Heinzmann}},
  \bibinfo{author}{\bibfnamefont{M.}~\bibnamefont{Drescher}}, \bibnamefont{and}
  \bibinfo{author}{\bibfnamefont{F.}~\bibnamefont{Krausz}},
  \bibinfo{journal}{Nature} \textbf{\bibinfo{volume}{427}},
  \bibinfo{pages}{817} (\bibinfo{year}{2004}).

\bibitem[{\citenamefont{Chew et~al.}(2012)\citenamefont{Chew, S{\"u}{\ss}mann,
  Sp{\"a}th, Wirth, Schmidt, Zherebtsov, Guggenmos, Oelsner, Weber, Kapaldo
  et~al.}}]{hhg1khz}
\bibinfo{author}{\bibfnamefont{S.~H.} \bibnamefont{Chew}},
  \bibinfo{author}{\bibfnamefont{F.}~\bibnamefont{S{\"u}{\ss}mann}},
  \bibinfo{author}{\bibfnamefont{C.}~\bibnamefont{Sp{\"a}th}},
  \bibinfo{author}{\bibfnamefont{A.}~\bibnamefont{Wirth}},
  \bibinfo{author}{\bibfnamefont{J.}~\bibnamefont{Schmidt}},
  \bibinfo{author}{\bibfnamefont{S.}~\bibnamefont{Zherebtsov}},
  \bibinfo{author}{\bibfnamefont{A.}~\bibnamefont{Guggenmos}},
  \bibinfo{author}{\bibfnamefont{A.}~\bibnamefont{Oelsner}},
  \bibinfo{author}{\bibfnamefont{N.}~\bibnamefont{Weber}},
  \bibinfo{author}{\bibfnamefont{J.}~\bibnamefont{Kapaldo}},
  \bibnamefont{et~al.}, \bibinfo{journal}{App. Phys. Lett.}
  \textbf{\bibinfo{volume}{100}}, \bibinfo{pages}{051904}
  (\bibinfo{year}{2012}).

\bibitem[{\citenamefont{Schultze et~al.}(2007)\citenamefont{Schultze,
  Goulielmakis, Uiberacker, Hofstetter, Kim, Kim, Krausz, and
  Kleineberg}}]{hhg3khz}
\bibinfo{author}{\bibfnamefont{M.}~\bibnamefont{Schultze}},
  \bibinfo{author}{\bibfnamefont{E.}~\bibnamefont{Goulielmakis}},
  \bibinfo{author}{\bibfnamefont{M.}~\bibnamefont{Uiberacker}},
  \bibinfo{author}{\bibfnamefont{M.}~\bibnamefont{Hofstetter}},
  \bibinfo{author}{\bibfnamefont{J.}~\bibnamefont{Kim}},
  \bibinfo{author}{\bibfnamefont{D.}~\bibnamefont{Kim}},
  \bibinfo{author}{\bibfnamefont{F.}~\bibnamefont{Krausz}}, \bibnamefont{and}
  \bibinfo{author}{\bibfnamefont{U.}~\bibnamefont{Kleineberg}},
  \bibinfo{journal}{New J. Phys.} \textbf{\bibinfo{volume}{9}},
  \bibinfo{pages}{243} (\bibinfo{year}{2007}).

\bibitem[{\citenamefont{Krebs et~al.}(2013)\citenamefont{Krebs, H{\"a}drich,
  Demmler, Rothhardt, Za{\"i}r, Chipperfield, Limpert, and
  T{\"u}nnermann}}]{amelle}
\bibinfo{author}{\bibfnamefont{M.}~\bibnamefont{Krebs}},
  \bibinfo{author}{\bibfnamefont{S.}~\bibnamefont{H{\"a}drich}},
  \bibinfo{author}{\bibfnamefont{S.}~\bibnamefont{Demmler}},
  \bibinfo{author}{\bibfnamefont{J.}~\bibnamefont{Rothhardt}},
  \bibinfo{author}{\bibfnamefont{A.}~\bibnamefont{Za{\"i}r}},
  \bibinfo{author}{\bibfnamefont{L.}~\bibnamefont{Chipperfield}},
  \bibinfo{author}{\bibfnamefont{J.}~\bibnamefont{Limpert}}, \bibnamefont{and}
  \bibinfo{author}{\bibfnamefont{A.}~\bibnamefont{T{\"u}nnermann}},
  \bibinfo{journal}{Nat. Phot.} \textbf{\bibinfo{volume}{7}},
  \bibinfo{pages}{555} (\bibinfo{year}{2013}).

\bibitem[{\citenamefont{Lein}(2007)}]{manfred_rev}
\bibinfo{author}{\bibfnamefont{M.}~\bibnamefont{Lein}}, \bibinfo{journal}{J.
  Phys. B} \textbf{\bibinfo{volume}{43}}, \bibinfo{pages}{R135}
  (\bibinfo{year}{2007}).

\bibitem[{\citenamefont{Baker et~al.}(2006)\citenamefont{Baker, Robinson,
  Haworth, Teng, Smith, Chiril{\u{a}}, Lein, Tisch, and Marangos}}]{pacer}
\bibinfo{author}{\bibfnamefont{S.}~\bibnamefont{Baker}},
  \bibinfo{author}{\bibfnamefont{J.~S.} \bibnamefont{Robinson}},
  \bibinfo{author}{\bibfnamefont{C.~A.} \bibnamefont{Haworth}},
  \bibinfo{author}{\bibfnamefont{H.}~\bibnamefont{Teng}},
  \bibinfo{author}{\bibfnamefont{R.~A.} \bibnamefont{Smith}},
  \bibinfo{author}{\bibfnamefont{C.~C.} \bibnamefont{Chiril{\u{a}}}},
  \bibinfo{author}{\bibfnamefont{M.}~\bibnamefont{Lein}},
  \bibinfo{author}{\bibfnamefont{J.~G.} \bibnamefont{Tisch}}, \bibnamefont{and}
  \bibinfo{author}{\bibfnamefont{J.~P.} \bibnamefont{Marangos}},
  \bibinfo{journal}{Science} \textbf{\bibinfo{volume}{312}},
  \bibinfo{pages}{424} (\bibinfo{year}{2006}).

\bibitem[{\citenamefont{Haessler et~al.}(2010)\citenamefont{Haessler, Caillat,
  Boutu, Giovanetti-Teixeira, Ruchon, Auguste, Diveki, Breger, Maquet, anb
  R.~Ta{\"{\i}}eb et~al.}}]{rabitt}
\bibinfo{author}{\bibfnamefont{S.}~\bibnamefont{Haessler}},
  \bibinfo{author}{\bibfnamefont{J.}~\bibnamefont{Caillat}},
  \bibinfo{author}{\bibfnamefont{W.}~\bibnamefont{Boutu}},
  \bibinfo{author}{\bibfnamefont{C.}~\bibnamefont{Giovanetti-Teixeira}},
  \bibinfo{author}{\bibfnamefont{T.}~\bibnamefont{Ruchon}},
  \bibinfo{author}{\bibfnamefont{T.}~\bibnamefont{Auguste}},
  \bibinfo{author}{\bibfnamefont{Z.}~\bibnamefont{Diveki}},
  \bibinfo{author}{\bibfnamefont{P.}~\bibnamefont{Breger}},
  \bibinfo{author}{\bibfnamefont{A.}~\bibnamefont{Maquet}},
  \bibinfo{author}{\bibfnamefont{B.~C.} \bibnamefont{anb R.~Ta{\"{\i}}eb}},
  \bibnamefont{et~al.}, \bibinfo{journal}{Nat. Phys.}
  \textbf{\bibinfo{volume}{6}}, \bibinfo{pages}{200} (\bibinfo{year}{2010}).

\bibitem[{\citenamefont{Smirnova
  et~al.}(2009{\natexlab{a}})\citenamefont{Smirnova, Mairesse, Patchkovskii,
  Dudovich, Villeneuve, Corkum, and Ivanov}}]{olga1}
\bibinfo{author}{\bibfnamefont{O.}~\bibnamefont{Smirnova}},
  \bibinfo{author}{\bibfnamefont{Y.}~\bibnamefont{Mairesse}},
  \bibinfo{author}{\bibfnamefont{S.}~\bibnamefont{Patchkovskii}},
  \bibinfo{author}{\bibfnamefont{N.}~\bibnamefont{Dudovich}},
  \bibinfo{author}{\bibfnamefont{D.}~\bibnamefont{Villeneuve}},
  \bibinfo{author}{\bibfnamefont{P.}~\bibnamefont{Corkum}}, \bibnamefont{and}
  \bibinfo{author}{\bibfnamefont{M.~Y.} \bibnamefont{Ivanov}},
  \bibinfo{journal}{Proc. Natl. Acad. Sci. USA} \textbf{\bibinfo{volume}{106}},
  \bibinfo{pages}{16556} (\bibinfo{year}{2009}{\natexlab{a}}).

\bibitem[{\citenamefont{Smirnova
  et~al.}(2009{\natexlab{b}})\citenamefont{Smirnova, Mairesse, Patchkovskii,
  Dudovich, Villeneuve, Corkum, and Ivanov}}]{olga2}
\bibinfo{author}{\bibfnamefont{O.}~\bibnamefont{Smirnova}},
  \bibinfo{author}{\bibfnamefont{Y.}~\bibnamefont{Mairesse}},
  \bibinfo{author}{\bibfnamefont{S.}~\bibnamefont{Patchkovskii}},
  \bibinfo{author}{\bibfnamefont{N.}~\bibnamefont{Dudovich}},
  \bibinfo{author}{\bibfnamefont{D.}~\bibnamefont{Villeneuve}},
  \bibinfo{author}{\bibfnamefont{P.}~\bibnamefont{Corkum}}, \bibnamefont{and}
  \bibinfo{author}{\bibfnamefont{M.~Y.} \bibnamefont{Ivanov}},
  \bibinfo{journal}{Nature (London)} \textbf{\bibinfo{volume}{460}},
  \bibinfo{pages}{972} (\bibinfo{year}{2009}{\natexlab{b}}).

\bibitem[{\citenamefont{Mairesse et~al.}(2003)\citenamefont{Mairesse, de~Bohan,
  Frasinski, Merdji, Dinu, Monchicourt, Breger, Kova\v{c}ev, Ta{\"i}eb,
  Carr{\'e} et~al.}}]{mairesse}
\bibinfo{author}{\bibfnamefont{Y.}~\bibnamefont{Mairesse}},
  \bibinfo{author}{\bibfnamefont{A.}~\bibnamefont{de~Bohan}},
  \bibinfo{author}{\bibfnamefont{L.~J.} \bibnamefont{Frasinski}},
  \bibinfo{author}{\bibfnamefont{H.}~\bibnamefont{Merdji}},
  \bibinfo{author}{\bibfnamefont{L.~C.} \bibnamefont{Dinu}},
  \bibinfo{author}{\bibfnamefont{P.}~\bibnamefont{Monchicourt}},
  \bibinfo{author}{\bibfnamefont{P.}~\bibnamefont{Breger}},
  \bibinfo{author}{\bibfnamefont{M.}~\bibnamefont{Kova\v{c}ev}},
  \bibinfo{author}{\bibfnamefont{R.}~\bibnamefont{Ta{\"i}eb}},
  \bibinfo{author}{\bibfnamefont{B.}~\bibnamefont{Carr{\'e}}},
  \bibnamefont{et~al.}, \bibinfo{journal}{Science}
  \textbf{\bibinfo{volume}{302}}, \bibinfo{pages}{1540} (\bibinfo{year}{2003}).

\bibitem[{\citenamefont{Power et~al.}(2010)\citenamefont{Power, March, Catoire,
  Sistrun, Krushelnick, Agostini, and DiMauro}}]{power}
\bibinfo{author}{\bibfnamefont{E.~P.} \bibnamefont{Power}},
  \bibinfo{author}{\bibfnamefont{A.~M.} \bibnamefont{March}},
  \bibinfo{author}{\bibfnamefont{F.}~\bibnamefont{Catoire}},
  \bibinfo{author}{\bibfnamefont{E.}~\bibnamefont{Sistrun}},
  \bibinfo{author}{\bibfnamefont{K.}~\bibnamefont{Krushelnick}},
  \bibinfo{author}{\bibfnamefont{P.}~\bibnamefont{Agostini}}, \bibnamefont{and}
  \bibinfo{author}{\bibfnamefont{L.~F.} \bibnamefont{DiMauro}},
  \bibinfo{journal}{Nat. Phot.} \textbf{\bibinfo{volume}{4}},
  \bibinfo{pages}{352} (\bibinfo{year}{2010}).

\bibitem[{\citenamefont{Kr{\"u}ger et~al.}(2011)\citenamefont{Kr{\"u}ger,
  Schenk, and Hommelhoff}}]{peter2011}
\bibinfo{author}{\bibfnamefont{M.}~\bibnamefont{Kr{\"u}ger}},
  \bibinfo{author}{\bibfnamefont{M.}~\bibnamefont{Schenk}}, \bibnamefont{and}
  \bibinfo{author}{\bibfnamefont{P.}~\bibnamefont{Hommelhoff}},
  \bibinfo{journal}{Nature} \textbf{\bibinfo{volume}{475}}, \bibinfo{pages}{78}
  (\bibinfo{year}{2011}).

\bibitem[{\citenamefont{Zherebtsov et~al.}(2011)\citenamefont{Zherebtsov,
  Fennel, Plenge, Antonsson, Znakovskaya, Wirth, Herrwerth, S{\"u}{\ss}mann,
  Peltz, Ahmad et~al.}}]{klingnature}
\bibinfo{author}{\bibfnamefont{S.}~\bibnamefont{Zherebtsov}},
  \bibinfo{author}{\bibfnamefont{T.}~\bibnamefont{Fennel}},
  \bibinfo{author}{\bibfnamefont{J.}~\bibnamefont{Plenge}},
  \bibinfo{author}{\bibfnamefont{E.}~\bibnamefont{Antonsson}},
  \bibinfo{author}{\bibfnamefont{I.}~\bibnamefont{Znakovskaya}},
  \bibinfo{author}{\bibfnamefont{A.}~\bibnamefont{Wirth}},
  \bibinfo{author}{\bibfnamefont{O.}~\bibnamefont{Herrwerth}},
  \bibinfo{author}{\bibfnamefont{F.}~\bibnamefont{S{\"u}{\ss}mann}},
  \bibinfo{author}{\bibfnamefont{C.}~\bibnamefont{Peltz}},
  \bibinfo{author}{\bibfnamefont{I.}~\bibnamefont{Ahmad}},
  \bibnamefont{et~al.}, \bibinfo{journal}{Nat. Physics}
  \textbf{\bibinfo{volume}{7}}, \bibinfo{pages}{656} (\bibinfo{year}{2011}).

\bibitem[{\citenamefont{Hutchison et~al.}(2012)\citenamefont{Hutchison, Ganeev,
  Witting, Frank, Okell, Tisch, and Marangos}}]{ganeev1}
\bibinfo{author}{\bibfnamefont{C.}~\bibnamefont{Hutchison}},
  \bibinfo{author}{\bibfnamefont{R.~A.} \bibnamefont{Ganeev}},
  \bibinfo{author}{\bibfnamefont{T.}~\bibnamefont{Witting}},
  \bibinfo{author}{\bibfnamefont{F.}~\bibnamefont{Frank}},
  \bibinfo{author}{\bibfnamefont{W.~A.} \bibnamefont{Okell}},
  \bibinfo{author}{\bibfnamefont{J.~W.~G.} \bibnamefont{Tisch}},
  \bibnamefont{and} \bibinfo{author}{\bibfnamefont{J.~P.}
  \bibnamefont{Marangos}}, \bibinfo{journal}{Opt. Lett.}
  \textbf{\bibinfo{volume}{37}}, \bibinfo{pages}{2064} (\bibinfo{year}{2012}).

\bibitem[{\citenamefont{Ganeev et~al.}(2012)\citenamefont{Ganeev, Witting,
  Hutchison, Frank, Redkin, Okell, Lei, Roschuk, Maier, Marangos
  et~al.}}]{ganeev2}
\bibinfo{author}{\bibfnamefont{R.~A.} \bibnamefont{Ganeev}},
  \bibinfo{author}{\bibfnamefont{T.}~\bibnamefont{Witting}},
  \bibinfo{author}{\bibfnamefont{C.}~\bibnamefont{Hutchison}},
  \bibinfo{author}{\bibfnamefont{F.}~\bibnamefont{Frank}},
  \bibinfo{author}{\bibfnamefont{P.~V.} \bibnamefont{Redkin}},
  \bibinfo{author}{\bibfnamefont{W.~A.} \bibnamefont{Okell}},
  \bibinfo{author}{\bibfnamefont{D.~Y.} \bibnamefont{Lei}},
  \bibinfo{author}{\bibfnamefont{T.}~\bibnamefont{Roschuk}},
  \bibinfo{author}{\bibfnamefont{S.~A.} \bibnamefont{Maier}},
  \bibinfo{author}{\bibfnamefont{J.~P.} \bibnamefont{Marangos}},
  \bibnamefont{et~al.}, \bibinfo{journal}{Phys. Rev. A}
  \textbf{\bibinfo{volume}{85}}, \bibinfo{pages}{015807}
  (\bibinfo{year}{2012}).

\bibitem[{\citenamefont{Kr{\"u}ger
  et~al.}(2012{\natexlab{a}})\citenamefont{Kr{\"u}ger, Schenk, F{\"o}rster, and
  Hommelhoff}}]{peter2012}
\bibinfo{author}{\bibfnamefont{M.}~\bibnamefont{Kr{\"u}ger}},
  \bibinfo{author}{\bibfnamefont{M.}~\bibnamefont{Schenk}},
  \bibinfo{author}{\bibfnamefont{M.}~\bibnamefont{F{\"o}rster}},
  \bibnamefont{and}
  \bibinfo{author}{\bibfnamefont{P.}~\bibnamefont{Hommelhoff}},
  \bibinfo{journal}{J. Phys. B} \textbf{\bibinfo{volume}{45}},
  \bibinfo{pages}{074006} (\bibinfo{year}{2012}{\natexlab{a}}).

\bibitem[{\citenamefont{Schenk et~al.}(2010)\citenamefont{Schenk, Kr{\"u}ger,
  and Hommelhoff}}]{peter2010}
\bibinfo{author}{\bibfnamefont{M.}~\bibnamefont{Schenk}},
  \bibinfo{author}{\bibfnamefont{M.}~\bibnamefont{Kr{\"u}ger}},
  \bibnamefont{and}
  \bibinfo{author}{\bibfnamefont{P.}~\bibnamefont{Hommelhoff}},
  \bibinfo{journal}{Phys. Rev. Lett.} \textbf{\bibinfo{volume}{105}},
  \bibinfo{pages}{257601} (\bibinfo{year}{2010}).

\bibitem[{\citenamefont{Herink et~al.}(2012)\citenamefont{Herink, Solli, Gulde,
  and Ropers}}]{herink}
\bibinfo{author}{\bibfnamefont{G.}~\bibnamefont{Herink}},
  \bibinfo{author}{\bibfnamefont{D.~R.} \bibnamefont{Solli}},
  \bibinfo{author}{\bibfnamefont{M.}~\bibnamefont{Gulde}}, \bibnamefont{and}
  \bibinfo{author}{\bibfnamefont{C.}~\bibnamefont{Ropers}},
  \bibinfo{journal}{Nature} \textbf{\bibinfo{volume}{483}},
  \bibinfo{pages}{190} (\bibinfo{year}{2012}).

\bibitem[{\citenamefont{Yalunin et~al.}(2011)\citenamefont{Yalunin, Gulde, and
  Ropers}}]{yalunin}
\bibinfo{author}{\bibfnamefont{S.~V.} \bibnamefont{Yalunin}},
  \bibinfo{author}{\bibfnamefont{M.}~\bibnamefont{Gulde}}, \bibnamefont{and}
  \bibinfo{author}{\bibfnamefont{C.}~\bibnamefont{Ropers}},
  \bibinfo{journal}{Phys. Rev. B} \textbf{\bibinfo{volume}{84}},
  \bibinfo{pages}{195426} (\bibinfo{year}{2011}).

\bibitem[{\citenamefont{Kr{\"u}ger
  et~al.}(2012{\natexlab{b}})\citenamefont{Kr{\"u}ger, Schenk, Hommelhoff,
  Watcher, Lemell, and Burgd{\"o}rfer}}]{michaelNJP}
\bibinfo{author}{\bibfnamefont{M.}~\bibnamefont{Kr{\"u}ger}},
  \bibinfo{author}{\bibfnamefont{M.}~\bibnamefont{Schenk}},
  \bibinfo{author}{\bibfnamefont{P.}~\bibnamefont{Hommelhoff}},
  \bibinfo{author}{\bibfnamefont{G.}~\bibnamefont{Watcher}},
  \bibinfo{author}{\bibfnamefont{C.}~\bibnamefont{Lemell}}, \bibnamefont{and}
  \bibinfo{author}{\bibfnamefont{J.}~\bibnamefont{Burgd{\"o}rfer}},
  \bibinfo{journal}{New J. Phys.} \textbf{\bibinfo{volume}{14}},
  \bibinfo{pages}{085019} (\bibinfo{year}{2012}{\natexlab{b}}).

\bibitem[{\citenamefont{Watcher et~al.}(2012)\citenamefont{Watcher, Lemell,
  Burgd{\"o}rfer, Schenk, Kr{\"u}ger, and Hommelhoff}}]{watcherPRB}
\bibinfo{author}{\bibfnamefont{G.}~\bibnamefont{Watcher}},
  \bibinfo{author}{\bibfnamefont{C.}~\bibnamefont{Lemell}},
  \bibinfo{author}{\bibfnamefont{J.}~\bibnamefont{Burgd{\"o}rfer}},
  \bibinfo{author}{\bibfnamefont{M.}~\bibnamefont{Schenk}},
  \bibinfo{author}{\bibfnamefont{M.}~\bibnamefont{Kr{\"u}ger}},
  \bibnamefont{and}
  \bibinfo{author}{\bibfnamefont{P.}~\bibnamefont{Hommelhoff}},
  \bibinfo{journal}{Phys. Rev. B} \textbf{\bibinfo{volume}{86}},
  \bibinfo{pages}{085019} (\bibinfo{year}{2012}).

\bibitem[{\citenamefont{Yalunin et~al.}(2013)\citenamefont{Yalunin, G, Solli,
  Kr{\"u}ger, Hommelhoff, Diehn, Munk, and Ropers}}]{ropersAdP}
\bibinfo{author}{\bibfnamefont{S.~V.} \bibnamefont{Yalunin}},
  \bibinfo{author}{\bibfnamefont{H.}~\bibnamefont{G}},
  \bibinfo{author}{\bibfnamefont{D.~R.} \bibnamefont{Solli}},
  \bibinfo{author}{\bibfnamefont{M.}~\bibnamefont{Kr{\"u}ger}},
  \bibinfo{author}{\bibfnamefont{P.}~\bibnamefont{Hommelhoff}},
  \bibinfo{author}{\bibfnamefont{M.}~\bibnamefont{Diehn}},
  \bibinfo{author}{\bibfnamefont{A.}~\bibnamefont{Munk}}, \bibnamefont{and}
  \bibinfo{author}{\bibfnamefont{C.}~\bibnamefont{Ropers}},
  \bibinfo{journal}{Ann. Phys.} \textbf{\bibinfo{volume}{525}},
  \bibinfo{pages}{L12} (\bibinfo{year}{2013}).

\bibitem[{\citenamefont{Ghimire et~al.}(2011)\citenamefont{Ghimire, DiChiara,
  Sistrunk, Agostini, DiMauro, and Reis}}]{ghimireexp}
\bibinfo{author}{\bibfnamefont{S.}~\bibnamefont{Ghimire}},
  \bibinfo{author}{\bibfnamefont{A.~D.} \bibnamefont{DiChiara}},
  \bibinfo{author}{\bibfnamefont{E.}~\bibnamefont{Sistrunk}},
  \bibinfo{author}{\bibfnamefont{P.}~\bibnamefont{Agostini}},
  \bibinfo{author}{\bibfnamefont{L.~F.} \bibnamefont{DiMauro}},
  \bibnamefont{and} \bibinfo{author}{\bibfnamefont{D.~A.} \bibnamefont{Reis}},
  \bibinfo{journal}{Nat. Phys.} \textbf{\bibinfo{volume}{7}},
  \bibinfo{pages}{138} (\bibinfo{year}{2011}).

\bibitem[{\citenamefont{Ghimire et~al.}(2012)\citenamefont{Ghimire, DiChiara,
  Sistrunk, Ndabashimiye, Szafruga, Mohammad, Agostini, DiMauro, and
  Reis}}]{ghimiretheory}
\bibinfo{author}{\bibfnamefont{S.}~\bibnamefont{Ghimire}},
  \bibinfo{author}{\bibfnamefont{A.~D.} \bibnamefont{DiChiara}},
  \bibinfo{author}{\bibfnamefont{E.}~\bibnamefont{Sistrunk}},
  \bibinfo{author}{\bibfnamefont{G.}~\bibnamefont{Ndabashimiye}},
  \bibinfo{author}{\bibfnamefont{U.~B.} \bibnamefont{Szafruga}},
  \bibinfo{author}{\bibfnamefont{A.}~\bibnamefont{Mohammad}},
  \bibinfo{author}{\bibfnamefont{P.}~\bibnamefont{Agostini}},
  \bibinfo{author}{\bibfnamefont{L.~F.} \bibnamefont{DiMauro}},
  \bibnamefont{and} \bibinfo{author}{\bibfnamefont{D.~A.} \bibnamefont{Reis}},
  \bibinfo{journal}{Phys. Rev. A} \textbf{\bibinfo{volume}{85}},
  \bibinfo{pages}{043836} (\bibinfo{year}{2012}).

\bibitem[{\citenamefont{Kim et~al.}(2008)\citenamefont{Kim, Jin, Kim, Park,
  Kim, and Kim}}]{kim}
\bibinfo{author}{\bibfnamefont{S.}~\bibnamefont{Kim}},
  \bibinfo{author}{\bibfnamefont{J.}~\bibnamefont{Jin}},
  \bibinfo{author}{\bibfnamefont{Y.-J.} \bibnamefont{Kim}},
  \bibinfo{author}{\bibfnamefont{I.-Y.} \bibnamefont{Park}},
  \bibinfo{author}{\bibfnamefont{Y.}~\bibnamefont{Kim}}, \bibnamefont{and}
  \bibinfo{author}{\bibfnamefont{S.-W.} \bibnamefont{Kim}},
  \bibinfo{journal}{Nature} \textbf{\bibinfo{volume}{453}},
  \bibinfo{pages}{757} (\bibinfo{year}{2008}).

\bibitem[{\citenamefont{M{\"u}hlschlegel
  et~al.}(2005)\citenamefont{M{\"u}hlschlegel, Eisler, Martin, Hecht, and
  Pohl}}]{muhl}
\bibinfo{author}{\bibfnamefont{P.}~\bibnamefont{M{\"u}hlschlegel}},
  \bibinfo{author}{\bibfnamefont{H.-J.} \bibnamefont{Eisler}},
  \bibinfo{author}{\bibfnamefont{O.~J.~F.} \bibnamefont{Martin}},
  \bibinfo{author}{\bibfnamefont{B.}~\bibnamefont{Hecht}}, \bibnamefont{and}
  \bibinfo{author}{\bibfnamefont{D.~W.} \bibnamefont{Pohl}},
  \bibinfo{journal}{Science} \textbf{\bibinfo{volume}{308}},
  \bibinfo{pages}{1607} (\bibinfo{year}{2005}).

\bibitem[{\citenamefont{Schuck et~al.}(2005)\citenamefont{Schuck, Fromm,
  Sundaramurthy, Kino, and Moerner}}]{schuck}
\bibinfo{author}{\bibfnamefont{P.~J.} \bibnamefont{Schuck}},
  \bibinfo{author}{\bibfnamefont{D.~P.} \bibnamefont{Fromm}},
  \bibinfo{author}{\bibfnamefont{A.}~\bibnamefont{Sundaramurthy}},
  \bibinfo{author}{\bibfnamefont{G.~S.} \bibnamefont{Kino}}, \bibnamefont{and}
  \bibinfo{author}{\bibfnamefont{W.~E.} \bibnamefont{Moerner}},
  \bibinfo{journal}{Phys. Rev. Lett.} \textbf{\bibinfo{volume}{94}},
  \bibinfo{pages}{017402} (\bibinfo{year}{2005}).

\bibitem[{\citenamefont{Husakou et~al.}(2011)\citenamefont{Husakou, Im, and
  Herrmann}}]{husakou}
\bibinfo{author}{\bibfnamefont{A.}~\bibnamefont{Husakou}},
  \bibinfo{author}{\bibfnamefont{S.-J.} \bibnamefont{Im}}, \bibnamefont{and}
  \bibinfo{author}{\bibfnamefont{J.}~\bibnamefont{Herrmann}},
  \bibinfo{journal}{Phys. Rev. A} \textbf{\bibinfo{volume}{83}},
  \bibinfo{pages}{043839} (\bibinfo{year}{2011}).

\bibitem[{\citenamefont{Yavuz et~al.}(2012)\citenamefont{Yavuz, Bleda, Altun,
  and Topcu}}]{yavuz}
\bibinfo{author}{\bibfnamefont{I.}~\bibnamefont{Yavuz}},
  \bibinfo{author}{\bibfnamefont{E.~A.} \bibnamefont{Bleda}},
  \bibinfo{author}{\bibfnamefont{Z.}~\bibnamefont{Altun}}, \bibnamefont{and}
  \bibinfo{author}{\bibfnamefont{T.}~\bibnamefont{Topcu}},
  \bibinfo{journal}{Phys. Rev. A} \textbf{\bibinfo{volume}{85}},
  \bibinfo{pages}{013416} (\bibinfo{year}{2012}).

\bibitem[{\citenamefont{Ciappina
  et~al.}(2012{\natexlab{a}})\citenamefont{Ciappina, Biegert, Quidant, and
  Lewenstein}}]{ciappi2012}
\bibinfo{author}{\bibfnamefont{M.~F.} \bibnamefont{Ciappina}},
  \bibinfo{author}{\bibfnamefont{J.}~\bibnamefont{Biegert}},
  \bibinfo{author}{\bibfnamefont{R.}~\bibnamefont{Quidant}}, \bibnamefont{and}
  \bibinfo{author}{\bibfnamefont{M.}~\bibnamefont{Lewenstein}},
  \bibinfo{journal}{Phys. Rev. A} \textbf{\bibinfo{volume}{85}},
  \bibinfo{pages}{033828} (\bibinfo{year}{2012}{\natexlab{a}}).

\bibitem[{\citenamefont{Shaaran et~al.}(2012)\citenamefont{Shaaran, Ciappina,
  and Lewenstein}}]{tahir2012}
\bibinfo{author}{\bibfnamefont{T.}~\bibnamefont{Shaaran}},
  \bibinfo{author}{\bibfnamefont{M.~F.} \bibnamefont{Ciappina}},
  \bibnamefont{and}
  \bibinfo{author}{\bibfnamefont{M.}~\bibnamefont{Lewenstein}},
  \bibinfo{journal}{Phys. Rev. A} \textbf{\bibinfo{volume}{86}},
  \bibinfo{pages}{023408} (\bibinfo{year}{2012}).

\bibitem[{\citenamefont{Ciappina
  et~al.}(2012{\natexlab{b}})\citenamefont{Ciappina, A\'{c}imovi\'{c}, Shaaran,
  Biegert, Quidant, and Lewenstein}}]{ciappi_opt}
\bibinfo{author}{\bibfnamefont{M.~F.} \bibnamefont{Ciappina}},
  \bibinfo{author}{\bibfnamefont{S.~S.} \bibnamefont{A\'{c}imovi\'{c}}},
  \bibinfo{author}{\bibfnamefont{T.}~\bibnamefont{Shaaran}},
  \bibinfo{author}{\bibfnamefont{J.}~\bibnamefont{Biegert}},
  \bibinfo{author}{\bibfnamefont{R.}~\bibnamefont{Quidant}}, \bibnamefont{and}
  \bibinfo{author}{\bibfnamefont{M.}~\bibnamefont{Lewenstein}},
  \bibinfo{journal}{Opt. Exp.} \textbf{\bibinfo{volume}{20}},
  \bibinfo{pages}{26261} (\bibinfo{year}{2012}{\natexlab{b}}).

\bibitem[{\citenamefont{Shaaran et~al.}(2013)\citenamefont{Shaaran, Ciappina,
  and Lewenstein}}]{ciappiAdP}
\bibinfo{author}{\bibfnamefont{T.}~\bibnamefont{Shaaran}},
  \bibinfo{author}{\bibfnamefont{M.~F.} \bibnamefont{Ciappina}},
  \bibnamefont{and}
  \bibinfo{author}{\bibfnamefont{M.}~\bibnamefont{Lewenstein}},
  \bibinfo{journal}{Ann. Phys.} \textbf{\bibinfo{volume}{525}},
  \bibinfo{pages}{97} (\bibinfo{year}{2013}).

\bibitem[{\citenamefont{Feti\'c et~al.}(2013)\citenamefont{Feti\'c,
  Kalajd\v{z}i\'c, and Milo\v{s}evi\'c}}]{miloAdP}
\bibinfo{author}{\bibfnamefont{B.}~\bibnamefont{Feti\'c}},
  \bibinfo{author}{\bibfnamefont{K.}~\bibnamefont{Kalajd\v{z}i\'c}},
  \bibnamefont{and} \bibinfo{author}{\bibfnamefont{D.~B.}
  \bibnamefont{Milo\v{s}evi\'c}}, \bibinfo{journal}{Ann. Phys.}
  \textbf{\bibinfo{volume}{525}}, \bibinfo{pages}{107} (\bibinfo{year}{2013}).

\bibitem[{\citenamefont{P{\'e}rez-Hern{\'a}ndez
  et~al.}(2013)\citenamefont{P{\'e}rez-Hern{\'a}ndez, Ciappina, Lewenstein,
  Roso, and Za{\"{\i}}r}}]{joseprl2013}
\bibinfo{author}{\bibfnamefont{J.~A.} \bibnamefont{P{\'e}rez-Hern{\'a}ndez}},
  \bibinfo{author}{\bibfnamefont{M.~F.} \bibnamefont{Ciappina}},
  \bibinfo{author}{\bibfnamefont{M.}~\bibnamefont{Lewenstein}},
  \bibinfo{author}{\bibfnamefont{L.}~\bibnamefont{Roso}}, \bibnamefont{and}
  \bibinfo{author}{\bibfnamefont{A.}~\bibnamefont{Za{\"{\i}}r}},
  \bibinfo{journal}{Phys. Rev. Lett.} \textbf{\bibinfo{volume}{110}},
  \bibinfo{pages}{053001} (\bibinfo{year}{2013}).

\bibitem[{\citenamefont{Yavuz}(2013)}]{yavuz2013}
\bibinfo{author}{\bibfnamefont{I.}~\bibnamefont{Yavuz}},
  \bibinfo{journal}{Phys. Rev. A} \textbf{\bibinfo{volume}{87}},
  \bibinfo{pages}{053815} (\bibinfo{year}{2013}).

\bibitem[{\citenamefont{Luo et~al.}(2013)\citenamefont{Luo, Li, Wang, Zhang,
  and Lu}}]{luo2013}
\bibinfo{author}{\bibfnamefont{J.}~\bibnamefont{Luo}},
  \bibinfo{author}{\bibfnamefont{Y.}~\bibnamefont{Li}},
  \bibinfo{author}{\bibfnamefont{Z.}~\bibnamefont{Wang}},
  \bibinfo{author}{\bibfnamefont{Q.}~\bibnamefont{Zhang}}, \bibnamefont{and}
  \bibinfo{author}{\bibfnamefont{P.}~\bibnamefont{Lu}}, \bibinfo{journal}{J.
  Phys. B} \textbf{\bibinfo{volume}{46}}, \bibinfo{pages}{145602}
  (\bibinfo{year}{2013}).

\bibitem[{\citenamefont{Sivis et~al.}(2012)\citenamefont{Sivis, Duwe, Abel, and
  Ropers}}]{ropersnat}
\bibinfo{author}{\bibfnamefont{M.}~\bibnamefont{Sivis}},
  \bibinfo{author}{\bibfnamefont{M.}~\bibnamefont{Duwe}},
  \bibinfo{author}{\bibfnamefont{B.}~\bibnamefont{Abel}}, \bibnamefont{and}
  \bibinfo{author}{\bibfnamefont{C.}~\bibnamefont{Ropers}},
  \bibinfo{journal}{Nature} \textbf{\bibinfo{volume}{485}}, \bibinfo{pages}{E1}
  (\bibinfo{year}{2012}).

\bibitem[{\citenamefont{Kim et~al.}(2012)\citenamefont{Kim, Jin, Kim, Park,
  Kim, and Kim}}]{Kimreply}
\bibinfo{author}{\bibfnamefont{S.}~\bibnamefont{Kim}},
  \bibinfo{author}{\bibfnamefont{J.}~\bibnamefont{Jin}},
  \bibinfo{author}{\bibfnamefont{Y.-J.} \bibnamefont{Kim}},
  \bibinfo{author}{\bibfnamefont{I.-Y.} \bibnamefont{Park}},
  \bibinfo{author}{\bibfnamefont{Y.}~\bibnamefont{Kim}}, \bibnamefont{and}
  \bibinfo{author}{\bibfnamefont{S.-W.} \bibnamefont{Kim}},
  \bibinfo{journal}{Nature} \textbf{\bibinfo{volume}{485}}, \bibinfo{pages}{E2}
  (\bibinfo{year}{2012}).

\bibitem[{\citenamefont{Sivis et~al.}(2013)\citenamefont{Sivis, Duwe, Abel, and
  Ropers}}]{sivis2013}
\bibinfo{author}{\bibfnamefont{M.}~\bibnamefont{Sivis}},
  \bibinfo{author}{\bibfnamefont{M.}~\bibnamefont{Duwe}},
  \bibinfo{author}{\bibfnamefont{B.}~\bibnamefont{Abel}}, \bibnamefont{and}
  \bibinfo{author}{\bibfnamefont{C.}~\bibnamefont{Ropers}},
  \bibinfo{journal}{Nat. Phys.} \textbf{\bibinfo{volume}{9}},
  \bibinfo{pages}{304} (\bibinfo{year}{2013}).

\bibitem[{\citenamefont{S{\"u}{\ss}mann and
  Kling}(2011{\natexlab{a}})}]{klingspie}
\bibinfo{author}{\bibfnamefont{F.}~\bibnamefont{S{\"u}{\ss}mann}}
  \bibnamefont{and} \bibinfo{author}{\bibfnamefont{M.~F.} \bibnamefont{Kling}},
  \bibinfo{journal}{Proc. of SPIE} \textbf{\bibinfo{volume}{8096}},
  \bibinfo{pages}{80961C} (\bibinfo{year}{2011}{\natexlab{a}}).

\bibitem[{\citenamefont{S{\"u}{\ss}mann and
  Kling}(2011{\natexlab{b}})}]{klingprb}
\bibinfo{author}{\bibfnamefont{F.}~\bibnamefont{S{\"u}{\ss}mann}}
  \bibnamefont{and} \bibinfo{author}{\bibfnamefont{M.~F.} \bibnamefont{Kling}},
  \bibinfo{journal}{Phys. Rev. B} \textbf{\bibinfo{volume}{84}},
  \bibinfo{pages}{121406(R)} (\bibinfo{year}{2011}{\natexlab{b}}).

\bibitem[{\citenamefont{Yang et~al.}(2013)\citenamefont{Yang, Scrinzi, Husakou,
  Li, Stebbings, S{\"u}{\ss}mann, Yu, Kim, R{\"u}hl, Herrmann
  et~al.}}]{lastkling}
\bibinfo{author}{\bibfnamefont{Y.-Y.} \bibnamefont{Yang}},
  \bibinfo{author}{\bibfnamefont{A.}~\bibnamefont{Scrinzi}},
  \bibinfo{author}{\bibfnamefont{A.}~\bibnamefont{Husakou}},
  \bibinfo{author}{\bibfnamefont{Q.-G.} \bibnamefont{Li}},
  \bibinfo{author}{\bibfnamefont{S.~L.} \bibnamefont{Stebbings}},
  \bibinfo{author}{\bibfnamefont{F.}~\bibnamefont{S{\"u}{\ss}mann}},
  \bibinfo{author}{\bibfnamefont{H.-J.} \bibnamefont{Yu}},
  \bibinfo{author}{\bibfnamefont{S.}~\bibnamefont{Kim}},
  \bibinfo{author}{\bibfnamefont{E.}~\bibnamefont{R{\"u}hl}},
  \bibinfo{author}{\bibfnamefont{J.}~\bibnamefont{Herrmann}},
  \bibnamefont{et~al.}, \bibinfo{journal}{Opt. Exp.}
  \textbf{\bibinfo{volume}{21}}, \bibinfo{pages}{2195} (\bibinfo{year}{2013}).

\bibitem[{\citenamefont{Hommelhoff
  et~al.}(2006{\natexlab{a}})\citenamefont{Hommelhoff, Sortais,
  Aghajani-Talesh, and Kasevich}}]{peter2006}
\bibinfo{author}{\bibfnamefont{P.}~\bibnamefont{Hommelhoff}},
  \bibinfo{author}{\bibfnamefont{Y.}~\bibnamefont{Sortais}},
  \bibinfo{author}{\bibfnamefont{A.}~\bibnamefont{Aghajani-Talesh}},
  \bibnamefont{and} \bibinfo{author}{\bibfnamefont{M.~A.}
  \bibnamefont{Kasevich}}, \bibinfo{journal}{Phys. Rev. Lett.}
  \textbf{\bibinfo{volume}{96}}, \bibinfo{pages}{077401}
  (\bibinfo{year}{2006}{\natexlab{a}}).

\bibitem[{\citenamefont{Hommelhoff
  et~al.}(2006{\natexlab{b}})\citenamefont{Hommelhoff, Kealhofer, and
  Kasevich}}]{peter2006a}
\bibinfo{author}{\bibfnamefont{P.}~\bibnamefont{Hommelhoff}},
  \bibinfo{author}{\bibfnamefont{C.}~\bibnamefont{Kealhofer}},
  \bibnamefont{and} \bibinfo{author}{\bibfnamefont{M.~A.}
  \bibnamefont{Kasevich}}, \bibinfo{journal}{Phys. Rev. Lett.}
  \textbf{\bibinfo{volume}{97}}, \bibinfo{pages}{247402}
  (\bibinfo{year}{2006}{\natexlab{b}}).

\bibitem[{\citenamefont{Ropers et~al.}(2007)\citenamefont{Ropers, Solli,
  Schulz, Lienau, and Elsaesser}}]{Ropers2007}
\bibinfo{author}{\bibfnamefont{C.}~\bibnamefont{Ropers}},
  \bibinfo{author}{\bibfnamefont{D.~R.} \bibnamefont{Solli}},
  \bibinfo{author}{\bibfnamefont{C.~P.} \bibnamefont{Schulz}},
  \bibinfo{author}{\bibfnamefont{C.}~\bibnamefont{Lienau}}, \bibnamefont{and}
  \bibinfo{author}{\bibfnamefont{T.}~\bibnamefont{Elsaesser}},
  \bibinfo{journal}{Phys. Rev. Lett.} \textbf{\bibinfo{volume}{98}},
  \bibinfo{pages}{043907} (\bibinfo{year}{2007}).

\bibitem[{\citenamefont{Barwick et~al.}(2007)\citenamefont{Barwick, Corder,
  Strohaber, Chandler-Smith, Uiterwaal, and Batelaan}}]{Barwick2007}
\bibinfo{author}{\bibfnamefont{B.}~\bibnamefont{Barwick}},
  \bibinfo{author}{\bibfnamefont{C.}~\bibnamefont{Corder}},
  \bibinfo{author}{\bibfnamefont{J.}~\bibnamefont{Strohaber}},
  \bibinfo{author}{\bibfnamefont{N.}~\bibnamefont{Chandler-Smith}},
  \bibinfo{author}{\bibfnamefont{C.}~\bibnamefont{Uiterwaal}},
  \bibnamefont{and} \bibinfo{author}{\bibfnamefont{H.}~\bibnamefont{Batelaan}},
  \bibinfo{journal}{New J. Phys.} \textbf{\bibinfo{volume}{9}},
  \bibinfo{pages}{142} (\bibinfo{year}{2007}).

\bibitem[{\citenamefont{Yanagisawa et~al.}(2009)\citenamefont{Yanagisawa,
  Hafner, Don{\'a}, Kl{\"o}ckner, Leuenberger, Greber, Hengsberger, and
  Osterwalder}}]{Yanagisawa2009}
\bibinfo{author}{\bibfnamefont{H.}~\bibnamefont{Yanagisawa}},
  \bibinfo{author}{\bibfnamefont{C.}~\bibnamefont{Hafner}},
  \bibinfo{author}{\bibfnamefont{P.}~\bibnamefont{Don{\'a}}},
  \bibinfo{author}{\bibfnamefont{M.}~\bibnamefont{Kl{\"o}ckner}},
  \bibinfo{author}{\bibfnamefont{D.}~\bibnamefont{Leuenberger}},
  \bibinfo{author}{\bibfnamefont{T.}~\bibnamefont{Greber}},
  \bibinfo{author}{\bibfnamefont{M.}~\bibnamefont{Hengsberger}},
  \bibnamefont{and}
  \bibinfo{author}{\bibfnamefont{J.}~\bibnamefont{Osterwalder}},
  \bibinfo{journal}{Phys. Rev. Lett.} \textbf{\bibinfo{volume}{103}},
  \bibinfo{pages}{257603} (\bibinfo{year}{2009}).

\bibitem[{\citenamefont{Bormann et~al.}(2010)\citenamefont{Bormann, Gulde,
  Weismann, Yalunin, and Ropers}}]{Bormann2010}
\bibinfo{author}{\bibfnamefont{R.}~\bibnamefont{Bormann}},
  \bibinfo{author}{\bibfnamefont{M.}~\bibnamefont{Gulde}},
  \bibinfo{author}{\bibfnamefont{A.}~\bibnamefont{Weismann}},
  \bibinfo{author}{\bibfnamefont{S.~V.} \bibnamefont{Yalunin}},
  \bibnamefont{and} \bibinfo{author}{\bibfnamefont{C.}~\bibnamefont{Ropers}},
  \bibinfo{journal}{Phys. Rev. Lett.} \textbf{\bibinfo{volume}{105}},
  \bibinfo{pages}{147601} (\bibinfo{year}{2010}).

\bibitem[{\citenamefont{Park et~al.}(2012)\citenamefont{Park, Piglosiewicz,
  Schmidt, Kollmann, Mascheck, and Lienau}}]{Park2012}
\bibinfo{author}{\bibfnamefont{D.~J.} \bibnamefont{Park}},
  \bibinfo{author}{\bibfnamefont{B.}~\bibnamefont{Piglosiewicz}},
  \bibinfo{author}{\bibfnamefont{S.}~\bibnamefont{Schmidt}},
  \bibinfo{author}{\bibfnamefont{H.}~\bibnamefont{Kollmann}},
  \bibinfo{author}{\bibfnamefont{M.}~\bibnamefont{Mascheck}}, \bibnamefont{and}
  \bibinfo{author}{\bibfnamefont{C.}~\bibnamefont{Lienau}},
  \bibinfo{journal}{Phys. Rev. Lett.} \textbf{\bibinfo{volume}{109}},
  \bibinfo{pages}{244803} (\bibinfo{year}{2012}).

\bibitem[{\citenamefont{Protopapas et~al.}(1997)\citenamefont{Protopapas,
  Keitel, and Knight}}]{keitel}
\bibinfo{author}{\bibfnamefont{M.}~\bibnamefont{Protopapas}},
  \bibinfo{author}{\bibfnamefont{C.~H.} \bibnamefont{Keitel}},
  \bibnamefont{and} \bibinfo{author}{\bibfnamefont{P.~L.}
  \bibnamefont{Knight}}, \bibinfo{journal}{Rep. Prog. Phys.}
  \textbf{\bibinfo{volume}{60}}, \bibinfo{pages}{389} (\bibinfo{year}{1997}).

\bibitem[{\citenamefont{Schafer and Kulander}(1997)}]{schafer}
\bibinfo{author}{\bibfnamefont{K.~J.} \bibnamefont{Schafer}} \bibnamefont{and}
  \bibinfo{author}{\bibfnamefont{K.~C.} \bibnamefont{Kulander}},
  \bibinfo{journal}{Phys. Rev. Lett.} \textbf{\bibinfo{volume}{78}},
  \bibinfo{pages}{638} (\bibinfo{year}{1997}).

\bibitem[{\citenamefont{Park et~al.}(2011)\citenamefont{Park, Kim, Choi, Lee,
  Kim, Kling, Stockman, and Kim}}]{funnel}
\bibinfo{author}{\bibfnamefont{I.-Y.} \bibnamefont{Park}},
  \bibinfo{author}{\bibfnamefont{S.}~\bibnamefont{Kim}},
  \bibinfo{author}{\bibfnamefont{J.}~\bibnamefont{Choi}},
  \bibinfo{author}{\bibfnamefont{D.-H.} \bibnamefont{Lee}},
  \bibinfo{author}{\bibfnamefont{Y.-J.} \bibnamefont{Kim}},
  \bibinfo{author}{\bibfnamefont{M.~F.} \bibnamefont{Kling}},
  \bibinfo{author}{\bibfnamefont{M.~I.} \bibnamefont{Stockman}},
  \bibnamefont{and} \bibinfo{author}{\bibfnamefont{S.-W.} \bibnamefont{Kim}},
  \bibinfo{journal}{Nat. Phot.} \textbf{\bibinfo{volume}{4}},
  \bibinfo{pages}{677} (\bibinfo{year}{2011}).

\bibitem[{\citenamefont{Sansone et~al.}(2009)\citenamefont{Sansone, Benedetti,
  Caumes, Stagira, Vozzi, and Nisoli}}]{tanos}
\bibinfo{author}{\bibfnamefont{G.}~\bibnamefont{Sansone}},
  \bibinfo{author}{\bibfnamefont{E.}~\bibnamefont{Benedetti}},
  \bibinfo{author}{\bibfnamefont{J.~P.} \bibnamefont{Caumes}},
  \bibinfo{author}{\bibfnamefont{S.}~\bibnamefont{Stagira}},
  \bibinfo{author}{\bibfnamefont{C.}~\bibnamefont{Vozzi}}, \bibnamefont{and}
  \bibinfo{author}{\bibfnamefont{M.}~\bibnamefont{Nisoli}},
  \bibinfo{journal}{Phys. Rev. A} \textbf{\bibinfo{volume}{80}},
  \bibinfo{pages}{063837} (\bibinfo{year}{2009}).

\bibitem[{\citenamefont{Corkum}(1993)}]{corkum}
\bibinfo{author}{\bibfnamefont{P.~B.} \bibnamefont{Corkum}},
  \bibinfo{journal}{Phys. Rev. Lett} \textbf{\bibinfo{volume}{71}},
  \bibinfo{pages}{1994} (\bibinfo{year}{1993}).

\bibitem[{\citenamefont{Lewenstein et~al.}(1994)\citenamefont{Lewenstein,
  Balcou, Ivanov, L'Huillier, and Corkum}}]{sfa}
\bibinfo{author}{\bibfnamefont{M.}~\bibnamefont{Lewenstein}},
  \bibinfo{author}{\bibfnamefont{P.}~\bibnamefont{Balcou}},
  \bibinfo{author}{\bibfnamefont{M.~Y.} \bibnamefont{Ivanov}},
  \bibinfo{author}{\bibfnamefont{A.}~\bibnamefont{L'Huillier}},
  \bibnamefont{and} \bibinfo{author}{\bibfnamefont{P.~B.}
  \bibnamefont{Corkum}}, \bibinfo{journal}{Phys. Rev. A}
  \textbf{\bibinfo{volume}{49}}, \bibinfo{pages}{2117} (\bibinfo{year}{1994}).

\bibitem[{\citenamefont{Thomas et~al.}(2012)\citenamefont{Thomas, Kr{\"u}ger,
  F{\"o}rster, Schenk, and Hommelhoff}}]{thomas}
\bibinfo{author}{\bibfnamefont{S.}~\bibnamefont{Thomas}},
  \bibinfo{author}{\bibfnamefont{M.}~\bibnamefont{Kr{\"u}ger}},
  \bibinfo{author}{\bibfnamefont{M.}~\bibnamefont{F{\"o}rster}},
  \bibinfo{author}{\bibfnamefont{M.}~\bibnamefont{Schenk}}, \bibnamefont{and}
  \bibinfo{author}{\bibfnamefont{P.}~\bibnamefont{Hommelhoff}},
  \bibinfo{journal}{arXiv:1209.5195}  (\bibinfo{year}{2012}).

\end{thebibliography}

%
\newpage
\begin{figure}[htb]
\centering
%\vspace{12cm}
%\hspace{-3cm}
\includegraphics[width=1.0\textwidth]{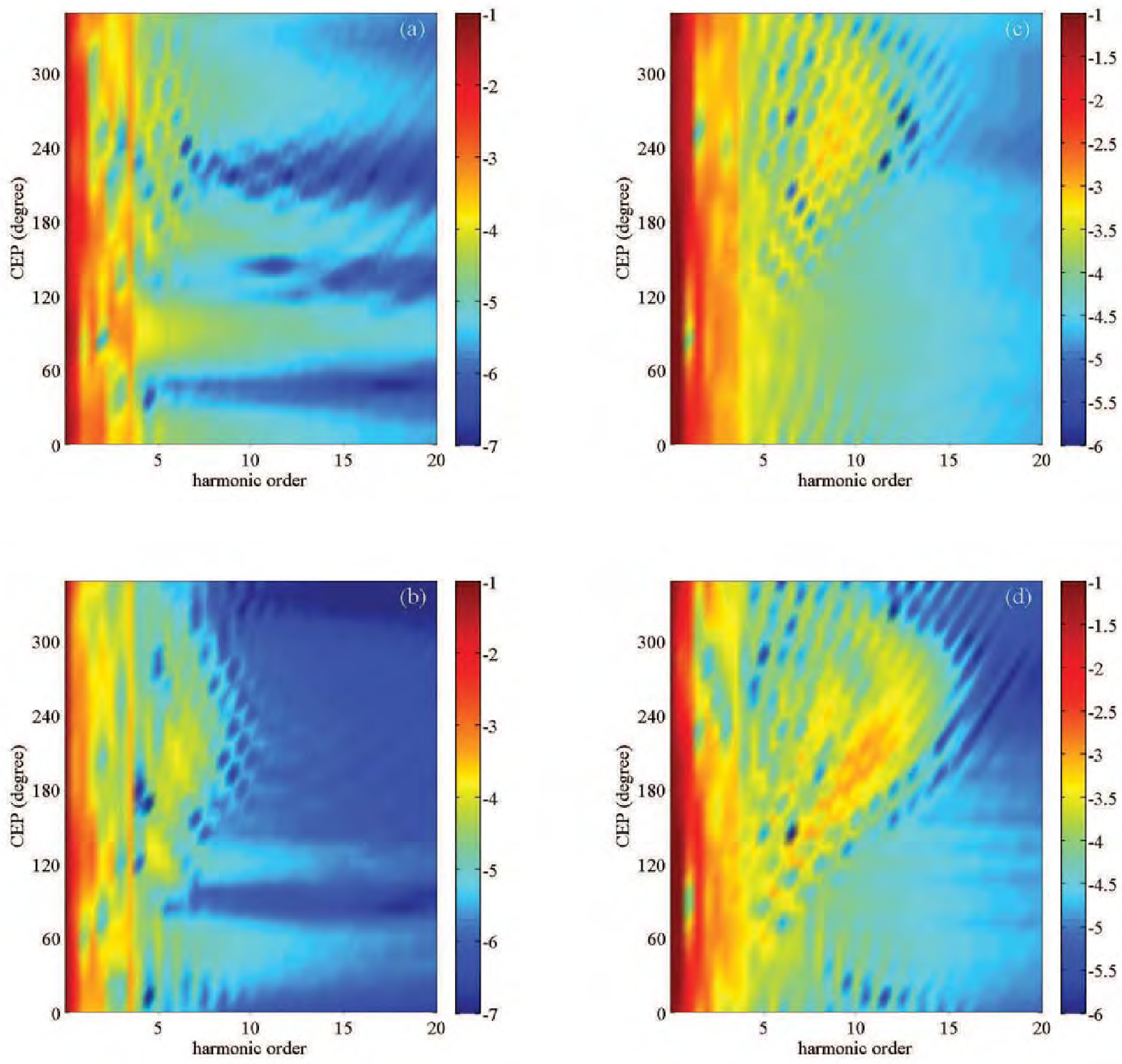}
%\resizebox{width=1\textwidth}{!}{\includegraphics{figure1.eps}}
%\vspace{-6cm}
\caption{(color online) Contour plots of the high-order harmonic spectra as a function harmonic order and carrier envelope phase (CEP) for a metal (Au) nanotip using a 2 fs FWHM laser pulse with a wavelength $\protect\lambda=800$ nm. Panel a) $E_0=10$ GV m$^{-1}$, $E_{dc}=-0.4$ GV m$^{-1}$; panel b) $E_0=10$ GV m$^{-1}$, $E_{dc}=+2$ GV m$^{-1}$; panel c) $E_0=20$ GV m$^{-1}$, $E_{dc}=-0.4$ GV m$^{-1}$ and panel d) $E_0=20$ GV m$^{-1}$, $E_{dc}=+2$ GV m$^{-1}$.}
\label{fig:figure1}
\end{figure}

\begin{figure}[htb]
\centering
%\vspace{12cm}
%\hspace{-3cm}
\includegraphics[width=1.0\textwidth]{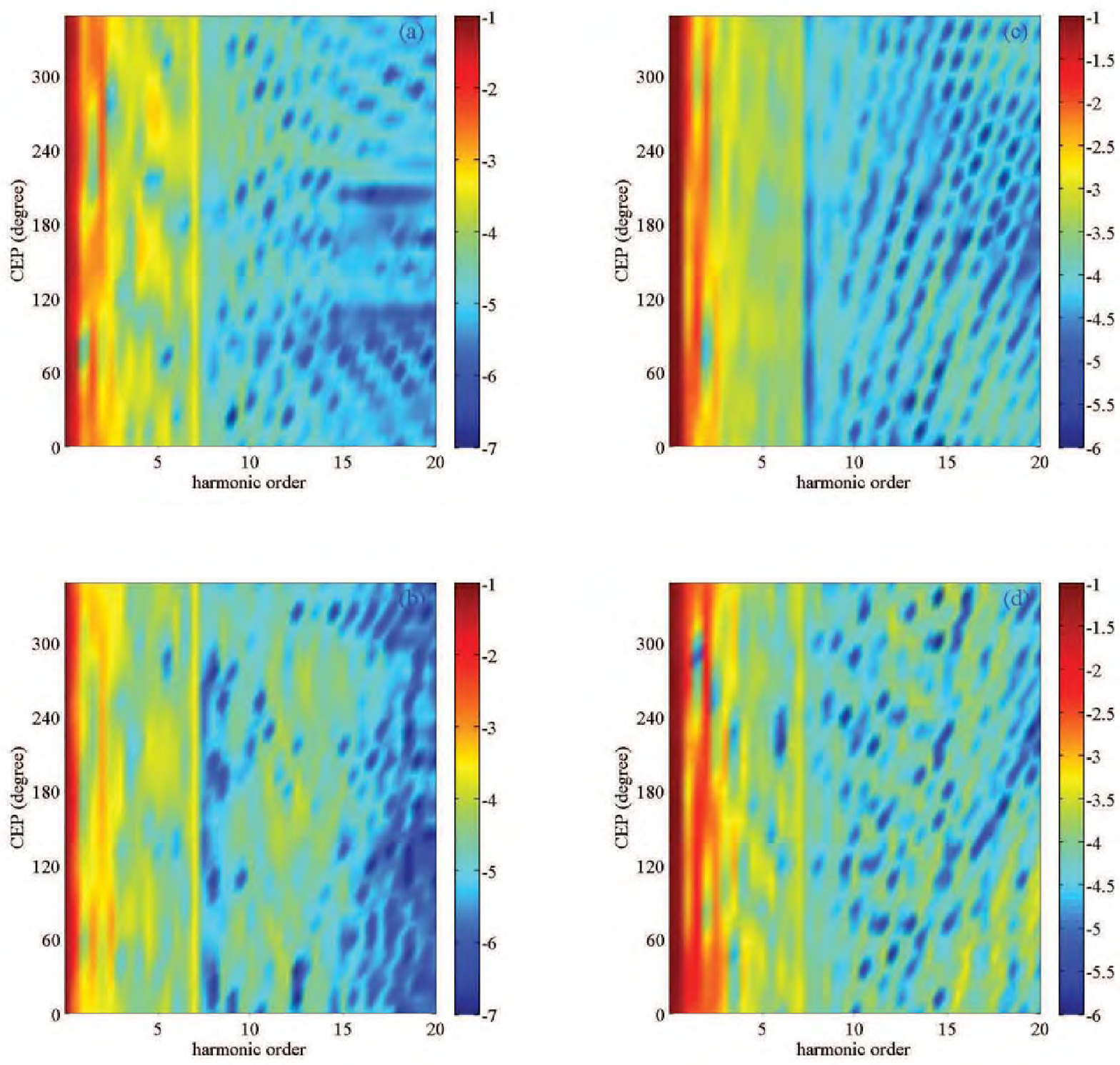}
%\vspace{-6cm}
\caption{(color online) Contour plots of the high-order harmonic spectra as a function harmonic order and carrier envelope phase (CEP) for a metal (Au) nanotip using a 4 fs FWHM laser pulse with a wavelength $\protect\lambda=800$ nm. Panel a) $E_0=10$ GV m$^{-1}$, $E_{dc}=-0.4$ GV m$^{-1}$; panel b) $E_0=10$ GV m$^{-1}$, $E_{dc}=+2$ GV m$^{-1}$; panel c) $E_0=20$ GV m$^{-1}$, $E_{dc}=-0.4$ GV m$^{-1}$ and panel d) $E_0=20$ GV m$^{-1}$, $E_{dc}=+2$ GV m$^{-1}$.}
\label{fig:figure2}
\end{figure}

%\begin{figure}[htb]
%\centering
%\vspace{12cm}
%\hspace{-3cm}
%\includegraphics[width=1\textwidth]{figure1}
%\vspace{-6cm}
%\caption{(color online) Contour plots of the high-order harmonic spectra as a function harmonic order and carrier envelope phase (CEP) for a metal (Au) nanotip using a 2 fs FWHM laser pulse with a wavelength $\protect\lambda=800$ nm. Panel a) $E_0=10$ GV m$^{-1}$, $E_{dc}=-0.4$ GV m$^{-1}$; panel b) $E_0=10$ GV m$^{-1}$, $E_{dc}=+2$ GV m$^{-1}$; panel c) $E_0=20$ GV m$^{-1}$, $E_{dc}=-0.4$ GV m$^{-1}$ and panel d) $E_0=20$ GV m$^{-1}$, $E_{dc}=+2$ GV m$^{-1}$.}
%\label{fig:figure1}
%\end{figure}

%\begin{figure}[htb]
%\centering
%\vspace{12cm}
%\hspace{-3cm}
%\includegraphics[width=1\textwidth]{figure2}
%\vspace{-6cm}
%\caption{(color online) Contour plots of the high-order harmonic spectra as a function harmonic order and carrier envelope phase (CEP) for a metal (Au) nanotip using a 4 fs FWHM laser pulse with a wavelength $\protect\lambda=800$ nm. Panel a) $E_0=10$ GV m$^{-1}$, $E_{dc}=-0.4$ GV m$^{-1}$; panel b) $E_0=10$ GV m$^{-1}$, $E_{dc}=+2$ GV m$^{-1}$; panel c) $E_0=20$ GV m$^{-1}$, $E_{dc}=-0.4$ GV m$^{-1}$ and panel d) $E_0=20$ GV m$^{-1}$, $E_{dc}=+2$ GV m$^{-1}$.}
%\label{fig:figure2}
%\end{figure}

\begin{figure}[htb]
\centering
%\hspace{-3cm}
\includegraphics[width=0.7\textwidth]{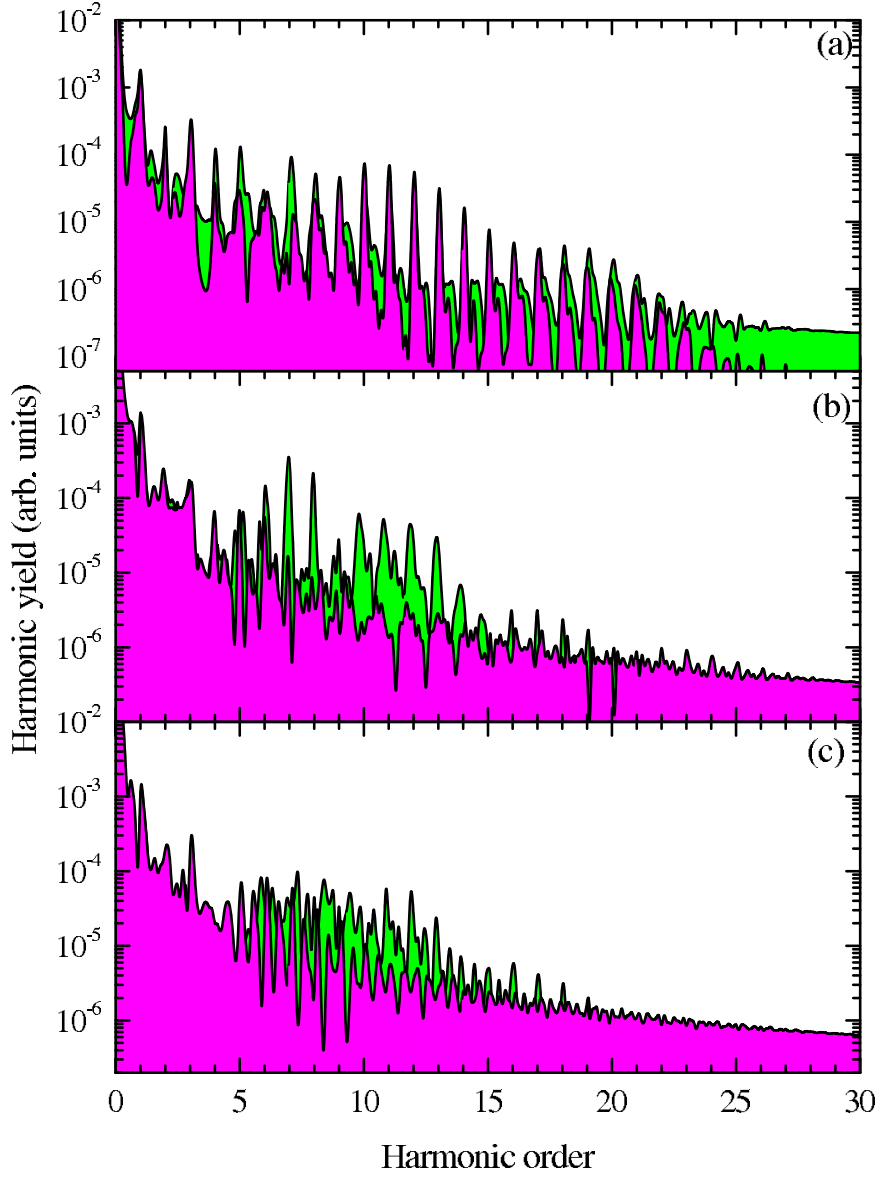}
%\vspace{-1cm}
\caption{(color online) HHG spectra as a function of harmonic order for a metal (Au) nanotip using a trapezoidal shaped laser pulse with 10 cycles of total time and a wavelength $\protect\lambda=685$ nm. Panel a) $E_0=10$ GV m$^{-1}$; panel b)  $E_0=15$ GV m$^{-1}$ and panel c)  $E_0=20$ GV m$^{-1}$. In all the panels magenta: $E_{dc}=-0.4$ GV m$^{-1}$, green: $E_{dc}=+2$ GV m$^{-1}$.}
\label{fig:figure3}
\end{figure}

\begin{figure}[htb]
\centering
%\hspace{-3cm}
\includegraphics[width=0.7\textwidth]{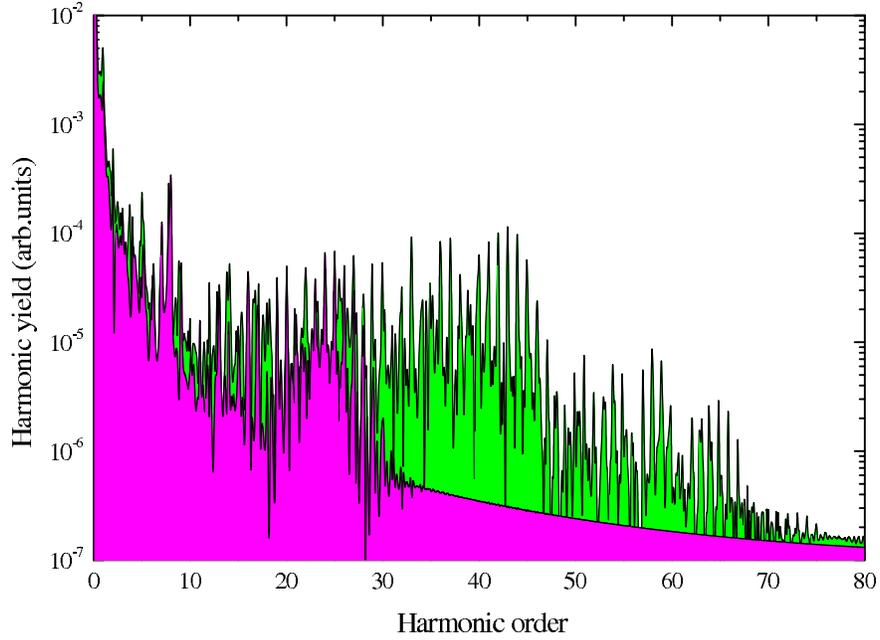}
%\vspace{-1cm}
\caption{(color online) HHG spectra as a function of harmonic order for a metal (Au) nanotip using a trapezoidal shaped laser pulse with 10 cycles of total time and a wavelength $\protect\lambda=1800$ nm and a peak laser electric field $E_0=10$ GV m$^{-1}$. Magenta $E_{dc}=-0.4$ GV m$^{-1}$, green $E_{dc}=+2$ GV m$^{-1}$.}
\label{fig:figure4}
\end{figure}

\end{document}